# THE $^{13}$C-POCKET STRUCTURE IN AGB MODELS: CONSTRAINTS FROM ZIRCONIUM ISOTOPE ABUNDANCES IN SINGLE MAINSTREAM SiC GRAINS

NAN LIU[1,2,3], ROBERTO GALLINO[4], SARA BISTERZO[4,5],

ANDREW M. DAVIS[1,2,6], MICHAEL R. SAVINA[2,3], MICHAEL J. PELLIN[1,2,3,6].

[1]Department of the Geophysical Sciences, The University of Chicago, Chicago, IL, 60637, USA; lnsmile@uchicago.edu;

[2]Chicago Center for Cosmochemistry, Chicago, IL 60637, USA;

[3]Materials Science Division, Argonne National Laboratory, Argonne, IL 60439, USA;

[4]Dipartimento di Fisica, Università di Torino, Torino I-10125, Italy;

[5]INAF−Osservatorio Astrofisico di Torino−Strada Osservatorio 20, Pino Torinese I-10025, Italy;

[6]Enrico Fermi Institute, The University of Chicago, Chicago, IL 60637, USA.

## ABSTRACT

We present postprocess AGB nucleosynthesis models with different $^{13}$C-pocket internal structures to better explain zirconium isotope measurements in mainstream **presolar SiC** grains by Nicolussi *et al.* (1997) and Barzyk *et al.* (2007). We show that higher-than-solar $^{92}$Zr/$^{94}$Zr ratios can be predicted by adopting a $^{13}$C-pocket with a flat $^{13}$C profile, instead of the previous decreasing-with-depth $^{13}$C profile. The improved agreement between grain data for zirconium isotopes and AGB models provides additional support for a recent proposal of a flat $^{13}$C profile based on barium isotopes in mainstream SiC grains by Liu *et al.* (2014).

## 1. INTRODUCTION

Studies of nuclear physics, stellar nucleosynthesis, astronomical observations and presolar grains play different roles in providing abundance predictions of nuclei synthesized in stars (Käppeler *et al.* 2011). While nucleosynthesis modeling of Asymptotic Giant Branch (AGB) stars is a method to study *s*-process nucleosynthesis, it requires accurate input data from nuclear physics, in particular, neutron capture probabilities expressed as the Maxwellian Averaged Cross-Section (MACS) of each nuclide. In addition, there exist problems in simulating formation of the $^{13}$C-pocket, a region rich in the major neutron source $^{13}$C, in AGB stellar models (e.g., Herwig 2005). AGB model predictions of *s*-process nuclide abundances therefore suffer from uncertainties in the shape and the size of the $^{13}$C-pocket (Bisterzo *et al.* 2014). **Mainstream presolar SiC** grains found in meteorites are produced in low-mass AGB stars with close-to-solar metallicity, and therefore carry *s*-process products from the convective envelope of parent stars during grain condensation (e.g., Gallino *et al.* 1990, Davis 2011). According to previous studies, AGB model predictions of *s*-process nuclide abundances can be constrained by precise isotopic studies of *s*-process elements in mainstream SiC grains (Lugaro *et al.* 2003, Barzyk *et al.* 2007, Liu *et al.* 2014).

Zirconium belongs to the first *s*-process peak and its isotopic abundances are sensitive to AGB stellar conditions. By adopting the MACSs recommended by Bao *et al.* (2000) and the stellar $\beta^-$ decay rates by Takahashi & Yokoi (1987), Lugaro *et al.* (2003) compared model predictions with SiC grain data from Nicolussi *et al.* (1997) and found a satisfactory agreement. Motivated by recently improved MACS data, Lugaro *et al.* (2014) reinvestigated zirconium isotope predictions using the AGB stellar models of Karakas (2010) with a range of initial stellar mass and metallicity, and found that $^{90,91,96}$Zr/$^{94}$Zr ratios in mainstream grains can be matched, but in the case of $^{92}$Zr/$^{94}$Zr, all predictions are significantly lower than the grain data.

Based on new barium isotope data obtained in acid-cleaned mainstream SiC grains, Liu *et al.* (2014) explored the effects of the internal structure of the $^{13}$C-pocket adopted in Torino AGB models on predictions for barium isotopes, and discovered that the predicted ratio of neutron-magic $^{138}$Ba to *s*-only $^{136}$Ba is extremely sensitive to the $^{13}$C profile within the $^{13}$C-pocket and the $^{13}$C-pocket mass. In order to reach a small group of mainstream grains with low $^{138}$Ba/$^{136}$Ba ratios, a pocket with a flat $^{13}$C profile and an initial mass less than $5.3 \times 10^{-4}\ M_{\odot}$ is needed. Inspired by the barium isotope study, we investigate effects of varying $^{13}$C-pocket structures on zirconium isotopes.

## 2. GRAIN DATA AND POSTPROCESS AGB MODELS

We use the zirconium isotope ratios in presolar SiC grains previously reported by Nicolussi *et al.* (1997) (renormalized to $^{94}$Zr by Davis *et al.* 1999) and Barzyk *et al.* (2007), and include four additional mainstream grains reported by Lugaro *et al.* (2003). All zirconium isotope data were obtained with the CHARISMA instrument using Resonance Ionization Mass Spectrometry (RIMS) at Argonne National Laboratory (Savina *et al.* 2003). Although the grains in the Nicolussi *et al.* study were not analyzed for carbon or silicon isotopes and could not be classified, we assume that the ones with *s*-process zirconium isotopic signatures are mainstream grains since > 90% of SiC grains are mainstream (Hoppe *et al.* 1994). Zirconium isotope data are expressed as δ-values with 2σ uncertainties, using $^{94}$Zr as the reference isotope. The δ-notation is defined as deviations in parts per thousand of isotope ratios in grains relative to those in terrestrial standards (e.g., $\delta(^{96}\mathrm{Zr}/^{94}\mathrm{Zr}) = [(^{96}\mathrm{Zr}/^{94}\mathrm{Zr})_{\mathrm{grain}}/(^{96}\mathrm{Zr}/^{94}\mathrm{Zr})_{\mathrm{standard}}-1)\times 1000]$).

An in-depth description of Torino postprocess AGB model calculations is given by Gallino *et al.* (1998) and updates are reported by Bisterzo *et al.* (2010). A simplified Three-zone $^{13}$C-pocket with a fixed mass of $9.37\times10^{-4}\ M_{\odot}$ was used **(hereafter Three-zone model with a decreasing $^{13}$C profile)**. In the Three-zone model, the pocket is subdivided into three different zones: Zone-I, -II & -III numbered in order of increasing stellar radius. The $^{13}$C mass fraction decreases from Zone-III to Zone-I with a fixed slope while remaining flat within each zone. A constant multiplication factor applied to the $^{13}$C mass fractions of each zone characterizes the $^{13}$C strength of the Three-zone $^{13}$C-pocket and is referred to as a *case*. The $^{13}$C mass fractions ($^{13}$C strengths) in D3−U2 cases are those of the ST (standard) case (in which the solar *s*-process pattern is well reproduced with a $0.5\ Z_{\odot}$ AGB star) divided (D) or multiplied (U) by the corresponding factors. For instance, the total $^{13}$C mass in the Three-zone $^{13}$C-pocket in the D1.5 case is

$$M(^{13}\mathrm{C})_{\mathrm{Three-zone}} = \frac{X(^{13}\mathrm{C})_{\mathrm{I}}}{1.5} \times M_{\mathrm{I}} + \frac{X(^{13}\mathrm{C})_{\mathrm{II}}}{1.5} \times M_{\mathrm{II}} + \frac{X(^{13}\mathrm{C})_{\mathrm{III}}}{1.5} \times M_{\mathrm{III}} \quad (1),$$

where $X(^{13}\mathrm{C})_{\mathrm{i}}$ and $M_{\mathrm{i}}$ are the $^{13}$C mass fraction and the mass of Zone-i, respectively, of the Three-zone $^{13}$C-pocket in the ST case.[1] For each zone, we also consider that some $^{14}$N, on the order of 1/30 of $^{13}$C, accompanies the $^{13}$C abundance and is produced by further proton capture via $^{13}$C$(p,\gamma)^{14}$N (Gallino *et al.* 1998). The Three-zone $^{13}$C-pocket mass is $9.37\times10^{-4}\ M_{\odot}$ ($M_{\rm I}+M_{\rm II}+M_{\rm III}$).

In Liu *et al.* (2014) and this study, we also explore a unique pocket with a flat $^{13}$C profile in which only the middle zone (Zone-II) is included (hereafter the Zone-II model). In addition, we explore the $^{13}$C-pocket mass as another parameter and vary its value in both the Three-zone and Zone-II model calculations. Taking the reference $^{13}$C-pocket masses in the Three-zone and Zone-II models as $9.37\times10^{-4}\ M_{\odot}$ and $5.3\times10^{-4}\ M_{\odot}$, respectively, we multiply (p) or divide (d) the $^{13}$C-pocket mass by the corresponding factor in each model's name. To summarize, three parameters, including the $^{13}$C mass fraction, $^{13}$C-pocket mass and $^{13}$C profile within the $^{13}$C-pocket, are considered in this study to characterize varying $^{13}$C-pockets for AGB model calculations. For instance, the total $^{13}$C masses in the Three-zone_d2 and the Zone-II_d2 $^{13}$C-pockets in the D1.5 case are

$$M(^{13}\mathrm{C})_{\mathrm{Three-zone_d2}} = \frac{X(^{13}\mathrm{C})_{\mathrm{I}}}{1.5}\times\frac{M_{\mathrm{I}}}{2} + \frac{X(^{13}\mathrm{C})_{\mathrm{II}}}{1.5}\times\frac{M_{\mathrm{II}}}{2} + \frac{X(^{13}\mathrm{C})_{\mathrm{III}}}{1.5}\times\frac{M_{\mathrm{III}}}{2} \quad (2)$$

$$M(^{13}\mathrm{C})_{\mathrm{Zone-II_d2}} = \frac{X(^{13}\mathrm{C})_{\mathrm{II}}}{1.5}\times\frac{M_{\mathrm{II}}}{2} \quad (3)$$, respectively.

As recently shown by Bisterzo *et al.* (2014), the solar *s*-process abundance distribution can also be well produced by using Zone-II AGB models coupled with Galactic Chemical Evolution (GCE). In this study, the solar abundances recommended by Lodders *et al.* (2009) are adopted and the solar metallicity is 0.0153 according to their Table 9.

Following the notation by Clayton (1968), MACS is defined as $\sigma^{i}_{\mathrm{MACS}} = \frac{<\sigma^{i} v>}{v_{\mathrm{T}}}$, where $\sigma^{i}$ is the $(n,\gamma)$ cross section of a nuclide $i$, $v$ the relative neutron velocity and $v_{\mathrm{T}}$ the mean thermal velocity. In the first approximation, MACSs are inversely proportional to $1/v_{\mathrm{T}}$ because of the general $1/v$ behavior of $\sigma^i$. We further define $\sigma_{code}{}^i$ as $\sigma^i_{code} = \frac{<\sigma^i v>}{v_T(30keV)} = \frac{\sigma^i_{MACS}\, v_T}{v_T(30keV)}$ (Lugaro *et al.* 2003). The product of $\sigma_{code}{}^i$ and $v_T$ is directly proportional to the rate of a given neutron capture reaction. Thus, $\sigma_{code}{}^i$ varies at different stellar temperatures if $\sigma_{MACS}{}^i$ deviates from $1/v_T$. This is the case for $^{91}$Zr and $^{92}$Zr. Their $\sigma_{MACS,}$ values deviate from $1/v_T$ by 30% from 8 keV to 23 keV as can be seen in Table 1. On the other hand, the $^{94}$Zr $\sigma_{MACS,}$ strictly follows the $1/v_T$ rule in this energy range. Thus, there is a dependence of $\delta(^{92}Zr/^{94}Zr)$ values on the two neutron sources, $^{22}$Ne and $^{13}$C, as will be discussed in Sections 3 & 4, respectively. For $^{90,91,92,93,94,96}$Zr, recent MACS measurements of Tagliente *et al.* (2008a, 2008b, 2010, 2011a, 2011b, 2013) are adopted. For the $^{95}$Zr MACS, there is an uncertainty of up to a factor of two (see KADoNiS[2] for details). The $^{95}$Zr MACS adopted in the Torino models are 50% of those recommended by KADoNiS, in agreement with Toukan & Käppeler (1990); even lower values (~30% of the KADoNiS values) are used in Lugaro *et al.* (2014).

[1] $M_{\rm I}= 4.0\times10^{-4}\ M_{\odot}$, $M_{\rm II}=5.3\times10^{-4}\ M_{\odot}$, $M_{\rm III}=7.5\times10^{-6}\ M_{\odot}$, $X(^{13}\mathrm{C})_{\rm I}= 3.2\times10^{-3}$, $X(^{13}\mathrm{C})_{\rm II}= 6.8\times10^{-3}$ and $X(^{13}\mathrm{C})_{\rm III}= 1.6\times10^{-2}$.

## 3. COMPARISON OF GRAIN DATA WITH TORINO THREE-ZONE AND ZONE-II AGB MODELS

We compare grain data with Three-zone and Zone-II models in Figure 1. Symbols are only plotted for thermal pulses (TPs) with envelope C/O > 1 for comparison with grain data, since this is when SiC is expected to condense based on thermodynamic equilibrium calculations (Lodders & Fegley 1995). The Three-zone and Zone-II predictions are shown as open and filled symbols, respectively. As shown in Figure 1, the Three-zone calculations with new MACSs match the grain data for $\delta(^{90}Zr/^{94}Zr)$ and $\delta(^{91}Zr/^{94}Zr)$. We also confirm the conclusion of Lugaro *et al.* (2003) **that the $^{89,90}$Sr and $^{91}$Y branch points have minimal impact on the final abundances of $^{90,91}$Zr[3]. Note that the adopted MACSs of $^{89,90}$Sr and $^{90,91}$Y are based on theoretical predictions (KADoNiS, Table II of Rauscher & Thielemann 2000).**

Previous studies of isotopic data on mainstream grains constrain the mass of their parent stars to 1.5−3 $M_\odot$ and the metallicity to close-to-solar metallicity (Barzyk *et al.* 2007). Lugaro *et*

---

[2]KADoNiS: Karlsruhe Astrophysical Database of Nucleosynthesis in Stars, website http://www.kadonis.org/, version v0.3.

[3]The recommended $^{89,90}$Sr and $^{91}$Y MACSs from Bao *et al.* (2000) are adopted in the Torino models used by both Lugaro *et al.* (2003) and this study.

*al.* (2014) concluded that the values of δ($^{92}$Zr/$^{94}$Zr) ≥ −50‰ observed in some grains cannot be reached by their new AGB models or by FRUITY models with various masses and metallicities within this range (Cristallo *et al.* 2011). We also compare Torino AGB models within this range of mass and metallicity with grain data in the three-isotope plots of δ($^{92}$Zr/$^{94}$Zr) versus δ($^{96}$Zr/$^{94}$Zr) in Figure 2. According to Figure 2, uncertainties in initial AGB stellar mass and metallicity cannot explain the mismatch of Torino AGB model predictions with the grains with δ($^{92}$Zr/$^{94}$Zr) ≥ −50‰, consistent with the conclusion from Lugaro *et al.* (2014).

As shown in Figure 2, model predictions for δ($^{92}$Zr/$^{94}$Zr) values are unaffected by variation of initial stellar mass and metallicity, while the δ($^{96}$Zr/$^{94}$Zr) prediction for the last TP increases with increasing initial stellar mass with its minimum remaining the same. This is because the final $^{96}$Zr production in AGB stars strongly depends on the $^{22}$Ne(*α,n*)$^{25}$Mg rate, which increases with increasing peak temperature at the bottom of the helium-burning zone. As the peak temperature increases with increasing core mass (Straniero *et al.* 2003), the $^{22}$Ne(*α,n*)$^{25}$Mg reaction operates more effectively in 3 $M_{\odot}$ AGB stars than in 1.5 $M_{\odot}$ AGB stars, resulting in higher $^{96}$Zr overproduction in the 3 $M_{\odot}$ case. On the other hand, *s*-process efficiency depends on the number of $^{13}$C nuclei per iron seed. As $^{13}$C is primary, produced by proton-capture on the freshly synthesized $^{12}$C in the helium intershell, the *s*-process efficiency depends linearly on the initial metallicity. For instance, Three-zone model predictions of $Z_{\odot}$ AGB stars in the U2 case are comparable to those of 0.5 $Z_{\odot}$ AGB stars in the ST case, because the number of iron seeds of $Z_{\odot}$ AGB stars is doubled with respect to that of 0.5 $Z_{\odot}$ AGB stars. **One-half-$Z_{\odot}$** AGB models, however, predict a longer carbon-rich phase and higher δ($^{96}$Zr/$^{94}$Zr) value for the last TP because 1) the convective envelope of 0.5 $Z_{\odot}$ AGB stars starts with less oxygen and therefore becomes carbon-rich after fewer pulses than that of $Z_{\odot}$ AGB stars; 2) the lower the metallicity, the higher the core mass and in turn, the higher the stellar temperature. Thus, the final $^{96}$Zr production is higher in 0.5 $Z_{\odot}$ AGB stars. To summarize, δ($^{92}$Zr/$^{94}$Zr) predictions are unaffected by uncertainties in the initial stellar mass and metallicity. We therefore use 2 $M_{\odot}$, 0.5 $Z_{\odot}$ AGB model as representative to investigate the effect of the $^{13}$C-pocket internal structure on $\delta(^{92}Zr/^{94}Zr)$ model predictions.

In addition, the production of $^{92}Zr$ and $^{94}Zr$ is unaffected by branching effects because different neutron capture paths flowing through branch points in this region all join at $^{92}Zr$. However, as noticed by Lugaro *et al.* (2003) and discussed in Section 2, the $^{92}Zr$ abundance is affected by the marginal activation of the $^{22}Ne(\alpha,n)^{25}Mg$ reaction during the TPs because the $^{92}Zr$ MACS deviates from $1/v_T$, whereas the $^{94}Zr$ MACS closely follows the $1/v_T$ rule. For reference, in Figure 3 $\delta(^{92}Zr/^{94}Zr)$ decreases by ~50‰ by varying the $^{22}Ne(\alpha,n)^{25}Mg$ rate from K94[4] to ½×K94 in the Zone-II_d2.5 model. This rate has been constrained to lie between ¼×K94 and K94 rate by Liu *et al.* (2014). Using the lowest rate (¼×K94) lowers the $\delta(^{92}Zr/^{94}Zr)$ Three-zone model predictions in Figure 1 by 100‰, which makes the disagreement with the grains with $\delta(^{92}Zr/^{94}Zr) \geq -50$‰ even more problematic. Interestingly, as shown in Figure 1, by switching to the Zone-II $^{13}C$-pocket, the 2 $M_\odot$, 0.5 $Z_\odot$ model predicts a wider range of $\delta(^{90,91,92}Zr/^{94}Zr)$ values. The cause of this change will be discussed in Section 4.1. While good agreement remains for $\delta(^{90,91}Zr/^{94}Zr)$ values, Zone-II models better match the grains with higher $\delta(^{92}Zr/^{94}Zr)$ values using the new $^{92}Zr$ MACS.

In addition to uncertainties in initial stellar mass and metallicity, AGB model predictions of $\delta(^{96}Zr/^{94}Zr)$ also suffer from uncertainties in the $^{22}Ne(\alpha,n)^{25}Mg$ rate and the $^{95}Zr$ MACS value. Due to its small MACS, the amount of $^{96}Zr$ destroyed by neutron capture during the interpulse period is negligible. Consequently, $^{96}Zr$ production depends linearly on the $^{95}Zr$ neutron capture rate (and therefore on the $^{95}Zr$ MACS value) at 23 keV during the marginal activation of the $^{22}Ne(\alpha,n)^{25}Mg$ reaction in the more advanced TPs, where the peak neutron density reaches ~$10^{10}$ $cm^{-3}$. Thus, model predictions for $\delta(^{96}Zr/^{94}Zr)$ are strongly affected by the $^{22}Ne(\alpha,n)^{25}Mg$ reaction. The higher the reaction rate, the higher the $^{95}Zr_n/^{95}Zr_{\beta-}$ ratio, and in turn, the higher the $^{96}Zr$ abundance ($^{95}Zr_n$ and $^{95}Zr_{\beta-}$ are the numbers of $^{95}Zr$ nuclei capturing a neutron or undergoing $\beta^-$ decay, respectively). As the highest constrained rate by Liu *et al.* (2014), K94, is adopted in all the model calculations in Figure 1, the possibility of a even higher $^{22}Ne(\alpha,n)^{25}Mg$ rate to explain the higher δ($^{96}$Zr/$^{94}$Zr) values is excluded. In addition, when adopting in the Torino models the lower $^{95}$Zr MACS given by Lugaro *et al.* (2014) (Section 2), the prediction for the last TP decreases by 50‰, making the match with the highest δ($^{96}$Zr/$^{94}$Zr) grain values even more problematic. Actually, the mismatch with theses two grains may be caused by different reasons: (1) solar zirconium contamination in these particular SiC grains; (2) the use of a higher $^{95}$Zr MACS; and/or (3) inefficient $^{96}$Zr destruction in $Z_{\odot}$ AGB stars with extremely low $^{13}$C strength (the D3 case in Figure 2). Although a larger $^{95}$Zr MACS is plausible considering its current uncertainty, it will result in the overproduction of $^{96}$Zr compared to its solar *s*-process abundance (Bisterzo *et al.* 2011). All Torino model predictions for δ($^{96}$Zr/$^{94}$Zr) are above −900‰ as shown in Figures 1 & 2, consistent with the model predictions by Lugaro *et al.* (2014) and FRUITY. Therefore, the model predictions mismatch several grains with even lower δ($^{96}$Zr/$^{94}$Zr) values. This mismatch needs to be confirmed in the future with more zirconium isotope data in mainstream **SiC** grains measured with higher precision. According to the discussion above, it is a difficult task for AGB models to yield precise predictions for δ($^{96}$Zr/$^{94}$Zr) values.

[4]K94: the lower limit of the $^{22}Ne(\alpha,n)^{25}Mg$ reaction rate recommended by Käppeler et al. (1994). Unless noted otherwise, the K94 rate is adopted in model calculations.

We focus on discussion of effects of the $^{13}$C-pocket internal structure on δ($^{92}$Zr/$^{94}$Zr) values in the following Section because the production of $^{92}$Zr and $^{94}$Zr is unaffected by branching effects. Although the Torino AGB model predictions of $^{90,91}$Zr final abundances are unaffected by the $^{89,90}$Sr and $^{91}$Y branch points by adopting MACSs recommended by Bao *et al.* (2000), model predictions for δ($^{90}$Zr/$^{94}$Zr) and δ($^{91}$Zr/$^{94}$Zr) values do suffer from uncertainties in the MACSs due to **a significant range among** theoretical predictions in the literature. Also, these theoretical predictions need to be confirmed by experimental measurements in the future.

## 4. DISCUSSION

### *4.1 Effects of $^{13}$C Profile and $^{13}$C-Pocket Mass*

AGB model predictions suffer from uncertainties in the $^{13}$C neutron source (Gallino *et al.* 1998), which depends on the adopted $^{13}$C profile and the $^{13}$C-pocket mass as described in Section 2. In *s*-process nucleosynthesis, the MACSs of most of the nuclei between magic numbers follow $1/v_T$ behavior and are therefore relatively insensitive to the internal structure of the $^{13}C$ neutron source within the $^{13}C$-pocket. However, $^{91,92}Zr$ are exceptions as their MACS values deviate from $1/v_T$ by 30% from 8 keV to 23 keV while the $^{94}Zr$ MACS closely follows $1/v_T$ (Table 1), which results in the dependence of $\delta(^{91,92}Zr/^{94}Zr)$ values on the $^{13}C$ profile in Figure 1[5]. As shown in Figure 1, the better agreement of Zone-II predictions with the grains with $\delta(^{92}Zr/^{94}Zr) \geq -50‰$ favors the Zone-II $^{13}C$-pocket over the Three-zone one.

AGB model predictions for $\delta(^{92}Zr/^{94}Zr)$ depend on the $^{13}C$ profile as shown above, however they are also a function of the $^{13}C$-pocket mass, which must be taken into account. Both Three-zone and Zone-II model predictions for $\delta(^{92}Zr/^{94}Zr)$ decrease with increasing $^{13}C$-pocket mass because $^{94}Zr$ has a smaller MACS than $^{92}Zr$, and can be accumulated more effectively with increasing $^{13}C$-pocket mass. The dependence of the Three-zone and Zone-II model predictions on the $^{13}C$-pocket mass is shown in Figure 4 as the D1.5 cases of Three-zone_d2 to Three-Zone_p2, and Zone-II_d2.5 to Zone-II_p2 models, where the $^{13}C$ mass fraction is held constant at the value corresponding to the D1.5 case while the $^{13}C$-pocket mass varies according to Equations (2) and (3), respectively. Model predictions at two TPs are shown: one for the 7$^{th}$ TP, which is the first carbon-rich pulse, and one for the 25$^{th}$ TP, which is the last pulse of the AGB phase. Figure 4 shows that Zone-II models produce higher $\delta(^{92}Zr/^{94}Zr)$ values for equivalent $^{13}C$-pocket masses, especially for the 25$^{th}$ TP (the last one), during which most grains could form due to higher mass loss per pulse and a higher C/O ratio. Thus, the effect of changing the $^{13}C$ profile is independent of the $^{13}C$-pocket mass and the $^{13}C$ mass fraction. The Three-zone_d2 model better matches the grains with higher $\delta(^{92}Zr/^{94}Zr)$ values than the other Three-zone models with higher $^{13}C$-pocket mass. All models that can reach the grains with $\delta(^{92}Zr/^{94}Zr) \geq -50‰$ are shown in Figure 3.

According to Liu *et al.* (2014) and this study, $\delta(^{92}Zr/^{94}Zr)$, $\delta(^{88}Sr/^{86}Sr)$ and $\delta(^{138}Ba/^{136}Ba)$ are tracers of the shape and the size of the $^{13}C$-pockets in AGB stars. In general, Zone-II models predict a wider range of values for these three tracers than the corresponding Three-zone models. In order to investigate the difference between the two sets of models, we ran 2 $M_{\odot}$, 0.5 $Z_{\odot}$ AGB models with 1) Zone-I, 2) Zone-II, and 3) Zone-I/II (a $^{13}$C-pocket with Zone-I and Zone-II) $^{13}$C-pockets in the D1.5 case as shown in Figure 5. First of all, we can see that model predictions of Zone-I/II are close to those of Three-zone in Figure 1 due to the fact that the mass of Zone-III is so low that its contribution to the *s*-process product abundances is negligible. The first-order explanation for the wider range of predictions by Zone-II models compared to the corresponding Three-zone models is that the $^{13}$C-pocket releases neutrons in radiative conditions resulting in distinct *s*-process isotopic signatures in each zone, which are then mixed together when the Zone-I/II (Three-zone) $^{13}$C-pocket is ingested by the next convective TP. Therefore, Zone-II $^{13}$C-pockets always predict a wider range of isotopic compositions than Three-zone ones as in the case of δ($^{88}$Sr/$^{86}$Sr) and δ($^{138}$Ba/$^{136}$Ba).

[5] In Figure 1 $\delta(^{90}Zr/^{94}Zr)$ also shows a slight dependence on the $^{13}C$ profile due to the small MACS value of neutron-magic $^{90}Zr$ (Liu *et al.* 2014).

However, as shown in Figure 5, the weighted averaged model predictions for δ($^{92}$Zr/$^{94}$Zr) and δ($^{138}$Ba//$^{136}$Ba) deviate from Zone-I/II predictions, which is, in particular, obvious in the case of δ($^{92}$Zr/$^{94}$Zr) as Zone-I/II predictions lie below both Zone-I and Zone-II predictions. Clearly, this is not simply the effect of averaging between two *s*-process isotopic signatures. Rather, this difference may result from the interplay between Zone-I and Zone-II during TPs. In other words, although the $^{13}$C($\alpha,n$)$^{16}$O reaction occurs radiatively during interpulses and *s*-process nucleosynthesis **processes** in the two zones are isolated from each other, the *s*-process products are mixed together during TPs when $^{22}$Ne($\alpha,n$)$^{25}$Mg occurs, whose efficiency increases with increasing TP numbers. Also, the mass involved in each TP is not constant and varies from pulse to pulse. Therefore, the difference between Zone-II and Three-zone model predictions for δ($^{92}$Zr/$^{94}$Zr) values is more likely due to the interplay between the two zones during TPs and consequent memories from the last TP during the subsequent interpulse phase. As there are 25 pulses in 2 $M_{\odot}$, 0.5 $Z_{\odot}$ AGB stars, the lower δ($^{92}$Zr/$^{94}$Zr) predictions by Three-zone models are caused by the cumulative effect. To summarize, although δ($^{88}$Sr/$^{86}$Sr), δ($^{92}$Zr/$^{94}$Zr), and δ($^{138}$Ba/$^{136}$Ba) model predictions are all sensitive to the shape and the size of the $^{13}$C-pocket, **the** **sensitivity of δ($^{92}$Zr/$^{94}$Zr) is caused by deviation of the $^{92}$Zr MACS from 1/$v_T$, while that of both δ($^{88}$Sr/$^{86}$Sr) and δ($^{138}$Ba/$^{136}$Ba) results from the fact that $^{88}$Sr and $^{138}$Ba are neutron-magic and therefore they act as bottlenecks in the *s*-process path due to their extremely small MACS values.**

Diverse $^{13}$C pockets could exist in parent AGB stars of mainstream grains, while the choice of the $^{22}$Ne($\alpha,n$)$^{25}$Mg rate depends on uncertainties in laboratory experiments. The derived constraints on the $^{22}$Ne($\alpha,n$)$^{25}$Mg rate using δ($^{134}$Ba/$^{136}$Ba) values in mainstream SiC grains by Liu *et al.* (2014) are in good agreement with recently recommended values by Jaeger *et al.* (2001) and Longland *et al.* (2012), $2.69 \times 10^{-11}\ cm^3\ mol^{-1}\ s^{-1}$ and $3.36\times10^{-11}\ cm^3\ mol^{-1}\ s^{-1}$, respectively. These values are within the range of ½×K94 to K94 rates. Therefore, the $^{22}$Ne($\alpha,n$)$^{25}$Mg rate in parent AGB stars of these mainstream grains must be between K94 and ½×K94. While the $^{13}$C pocket structure and mass are free parameters in most AGB models, various $^{13}$C-pockets likely exist in different AGB stars because of different degrees of mixing and shaping induced by the physical process(es) responsible for formation of the $^{13}$C-pocket. For instance, in addition to mixing of hydrogen into the helium intershell by diverse overshoot diffusion mechanisms (e.g., Herwig et al. 1997, Cristallo et al. 2009, 2011) and/or gravity waves (Denissenkov & Tout, 2003) to form the $^{13}$C-pocket at the bottom of the convective envelope after a Third Dredge-up (TDU), the Goldreich-Schubert-Fricke (GSF) instability could further shape the $^{13}$C-pocket during interpulses in rotating AGB stars with close-to-solar metallicity (e.g., Piersanti *et al.* 2013). Thus, although the Zone-II_d2.5 model can explain most of the grain data for zirconium isotopes, we cannot exclude the possibility of diverse $^{13}$C-pockets in parent AGB stars because, for instance, the Three-zone model still explains most of the grains with δ($^{92}$Zr/$^{94}$Zr) ≤ −50‰. Further investigation of this possibility requires comparison of model predictions with grain data using at least two tracers of the $^{13}$C-pocket, e.g., correlated δ($^{92}$Zr/$^{94}$Zr) and δ($^{138}$Ba/$^{136}$Ba) values.

### *4.2 Effects of $^{92}$Zr and $^{94}$Zr MACS Uncertainties*

We evaluated the effects of 1σ uncertainties of the $^{92}$Zr and $^{94}$Zr MACSs in Table 1 on $\delta(^{92}Zr/^{94}Zr)$ predictions. For Three-zone calculations, the maximum $\delta(^{92}Zr/^{94}Zr)$ in the ST case decreases by 50‰ by adopting the upper limit (UL) $^{92}Zr$ MACS or the lower limit (LL) $^{94}Zr$ MACS; it increases by 60‰ by adopting the LL $^{92}Zr$ MACS or the UL $^{94}Zr$ MACS. Due to the MACS uncertainties, we are not able to exclude the possibility of a lower $^{92}Zr$ MACS and/or a higher $^{94}Zr$ MACS to explain the mismatch by Lugaro *et al.* (2014), and by Three-zone models for $\delta(^{92}Zr/^{94}Zr) \geq -50$‰. For instance, by adopting both the LL $^{92}Zr$ MACS and the UL $^{94}Zr$ MACS, the Three-zone model predictions in Figure 1 will be shifted by ~100‰ to higher $\delta(^{92}Zr/^{94}Zr)$ values and a better agreement could be obtained with the grain data. Therefore, higher precision measurements of the $^{92}Zr$ and $^{94}Zr$ MACSs ($2\sigma < 5\%$) are needed to determine whether Zone-II models are needed.

On the other hand, thanks to the well-determined MACSs of $^{136}Ba$ and $^{138}Ba$, and the barium isotopes in acid-cleaned mainstream grains, Liu *et al.* (2014) were able to demonstrate that the ST cases in Zone-II_d2.5 to Zone-II models are required to explain $\delta(^{138}Ba/^{136}Ba)$ values < −400‰ in a minor group of mainstream grains. The fact that the Zone-II models with lower $^{13}C$-pocket mass can better reach the grains with $\delta(^{92}Zr/^{94}Zr) \geq -50$‰ than the Three-zone models strongly supports lower-mass $^{13}C$-pockets with flatter $^{13}C$ profiles in the parent stars of mainstream grains. On the other hand, although Zone-II models with lower $^{13}C$-pocket mass can explain the whole range of $\delta(^{138}Ba/^{136}Ba)$ values observed by Liu *et al.* (2014), we cannot exclude the possibility of varying $^{13}C$-pockets existing in their parent stars, as Zone-II models are only needed to explain $\delta(^{138}Ba/^{136}Ba) < -400$‰.

## 5. CONCLUSION

The better agreement of zirconium isotope data, $\delta(^{92}Zr/^{94}Zr) \geq -50$‰ in particular, in mainstream SiC gains with the Torino Zone-II AGB model calculations strongly supports our previous proposal of a Zone-II $^{13}C$-pocket with the $^{13}C$-pocket mass less than $5.3 \times 10^{-4} M_{\odot}$ to explain mainstream grains with $\delta(^{138}Ba/^{136}Ba) < -400$‰ (Liu *et al.* 2014). All of the Zone-II models explored so far predict higher $\delta(^{92}Zr/^{94}Zr)$ values than their corresponding Three-zone models. A general trend of decreasing $\delta(^{92}Zr/^{94}Zr)$ values with increasing $^{13}C$-pocket mass is observed for both the Three-zone and Zone-II models, which is caused by the higher $^{94}$Zr overproduction. Due to uncertainties in zirconium isotope MACSs, we cannot completely exclude the Three-zone models in explaining existing grain data. MACS measurements of $^{92,94}$Zr with smaller uncertainties are needed to clarify this point. Correlated zirconium and barium isotope measurements in acid-cleaned mainstream grains free from contamination with material with solar isotopic ratios are needed to investigate the correlation between $\delta(^{92}Zr/^{94}Zr)$ and $\delta(^{138}Ba/^{136}Ba)$. Comparison of such acid-cleaned grain data with AGB model predictions with reduced uncertainties will eventually shed light onto the longstanding puzzle of the profile and the mass of the $^{13}$C-pocket and answer the question: what fraction of the parent AGB stars has a flatter $^{13}$C-pocket than the previous decreasing $^{13}$C-pocket?

Acknowledgements: We thank the anonymous referee for a careful and constructive reading of the manuscript. Part of the Torino model numerical calculations has been sustained by the B2FH Association (http://www.b2fh.org/). NL acknowledges the NASA Earth and Space Sciences Fellowship Program (NNX11AN63H) for support. This work was also supported by the NASA Cosmochemistry program through grant NNX09AG39G (to AMD). SB acknowledges financial support from the Joint Institute for Nuclear Astrophysics (JINA, University of Notre Dame, USA) and from Karlsruhe Institute of Technology (KIT, Karlsruhe, Germany).

FIGURE CAPTIONS

Figure 1 Three-isotope plots of $\delta(^{90}Zr/^{94}Zr)$, $\delta(^{91}Zr/^{94}Zr)$ and $\delta(^{92}Zr/^{94}Zr)$ versus $\delta(^{96}Zr/^{94}Zr)$. The grains (data sources in Section 2, 2σ uncertainty) are compared to Three-zone (left, open symbols) and Zone-II (right, filled symbols) AGB model predictions for a 2 $M_{\odot}$, 0.5 $Z_{\odot}$ AGB star with a range of $^{13}C$ strengths. The entire evolution of the AGB envelope composition is shown, but symbols are plotted only when C > O. Dotted lines represent solar zirconium isotope ratios.

Figure 2 Three-isotope plots of $\delta(^{92}Zr/^{94}Zr)$ versus $\delta(^{96}Zr/^{94}Zr)$. Grain data from Figure 1 are compared to Three-zone models for AGB stars with a range of mass and metallicity.

Figure 3 Three-isotope plots of $\delta(^{92}Zr/^{94}Zr)$ versus $\delta(^{96}Zr/^{94}Zr)$. The same grain data as in Figures 1 & 2 are compared to Three-zone (open symbols) and Zone-II (filled symbols) models for a 2 $M_{\odot}$, 0.5 $Z_{\odot}$ AGB star. Model descriptions are given in Section 2.

Figure 4 The $\delta(^{92}Zr/^{94}Zr)$ model predictions for the D1.5 case of a 2 $M_{\odot}$, 0.5 $Z_{\odot}$ AGB star at the 7$^{th}$ TP (when the convective envelope becomes carbon-rich) and 25$^{th}$ TP (the last TP during the AGB phase) are plotted against the $^{13}C$-pocket mass for Three-Zone_d2 to Three-Zone_p2, and Zone-II_d2.5 to Zone-II_p2 models.

Figure 5 Torino AGB model predictions with various $^{13}C$-pocket internal structures in the D1.5 case for a 2 $M_{\odot}$, 0.5 $Z_{\odot}$ AGB star are shown in the three-isotope plots of $\delta(^{92}Zr/^{94}Zr)$ versus $\delta(^{96}Zr/^{94}Zr)$, $\delta(^{88}Sr/^{86}Sr)$ versus $\delta(^{84}Sr/^{86}Sr)$, and $\delta(^{138}Ba/^{136}Ba)$ versus $\delta(^{135}Ba/^{136}Ba)$.

Table 1 Zirconium Neutron Capture Cross Sections, $\sigma^i_{code} = \frac{<\sigma^i v>}{v_{T(30keV)}} = \frac{\sigma^i_{MACS} v_T}{v_{T(30keV)}}$ (mbarn)

| Isotope | LDG03[a] | | | | This work[b] | | | |
|---|---|---|---|---|---|---|---|---|
| | 8 keV | 23 keV | 30 keV | 1σ err (30 keV) | 8 keV | 23 keV | 30 keV | 1σ err (30 keV) |
| $^{90}Zr$ | 18.9 | 20.8 | 21.0 | 9.5% | 18.1 | 18.8 | 19.3 | 4.7% |
| $^{91}Zr$ | 90.5 | 64.7 | 60.0 | 13.3% | 87.3 | 67.4 | 63.0 | 6.3% |
| $^{92}Zr$ | 47.2 | 34.6 | 33.0 | 12.1% | 50.1 | 38.5 | 38.0 | 8.0% |
| $^{93}Zr$ | 128.0 | 103.0 | 95.0 | 10.5% | 127.9 | 102.8 | 95.0 | 9.4% |
| $^{94}Zr$ | 30.1 | 26.6 | 26.0 | 3.8% | 30.2 | 29.8 | 30.5 | 6.5% |
| $^{95}Zr^*$ | 111.1 | 85.2 | 79.0 | 15.0% | 55.5 | 42.6 | 39.5 | 15.0% |
| $^{96}Zr$ | 18.1 | 11.1 | 10.7 | 4.7% | 17.2 | 10.9 | 10.3 | 6.0% |

Notes: [a] Lugaro *et al.* (2003) and references therein;

[b] See text for references;

[*] The 1σ uncertainty is only based on the theoretical estimate from Bao *et al.* (2000).

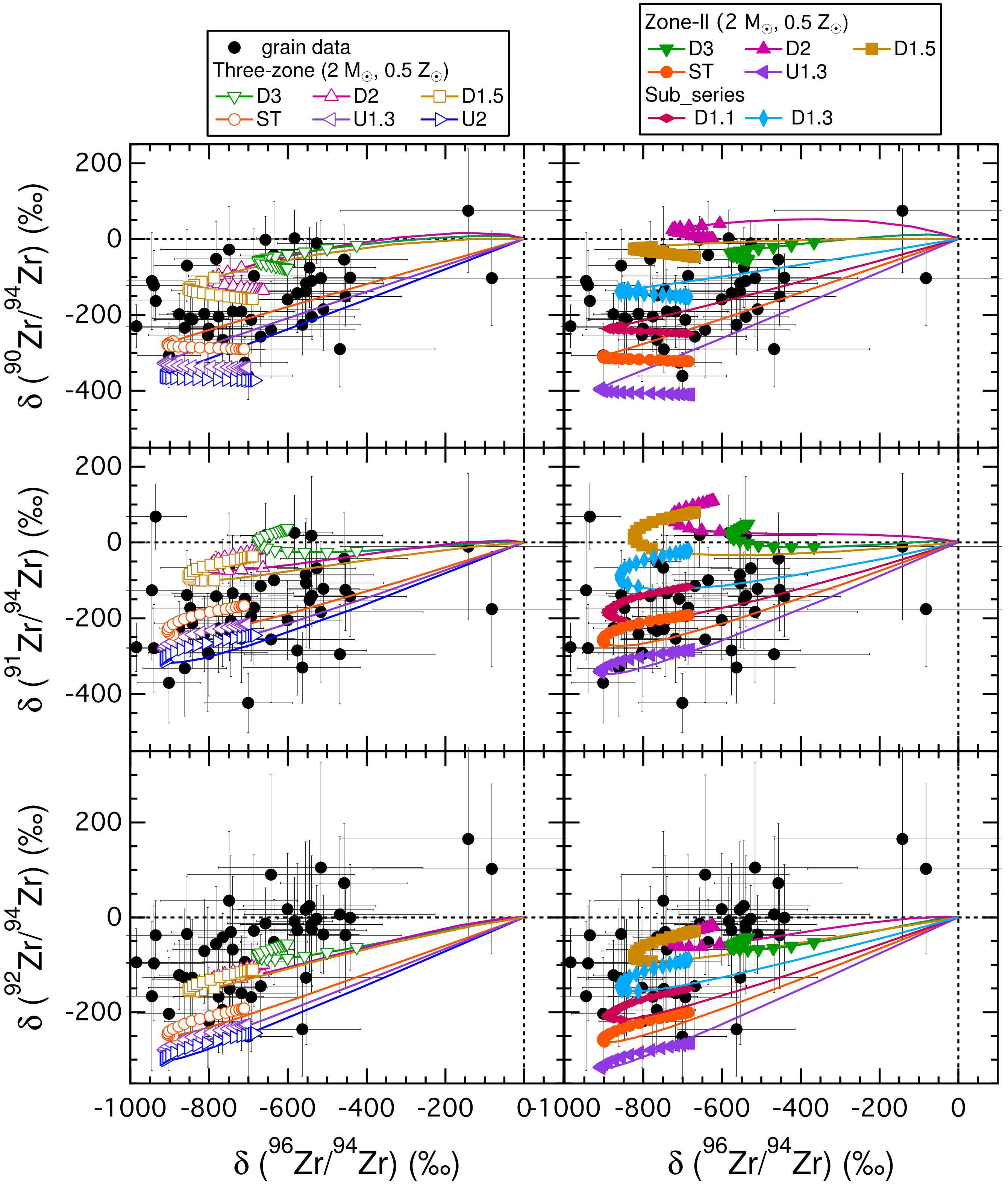

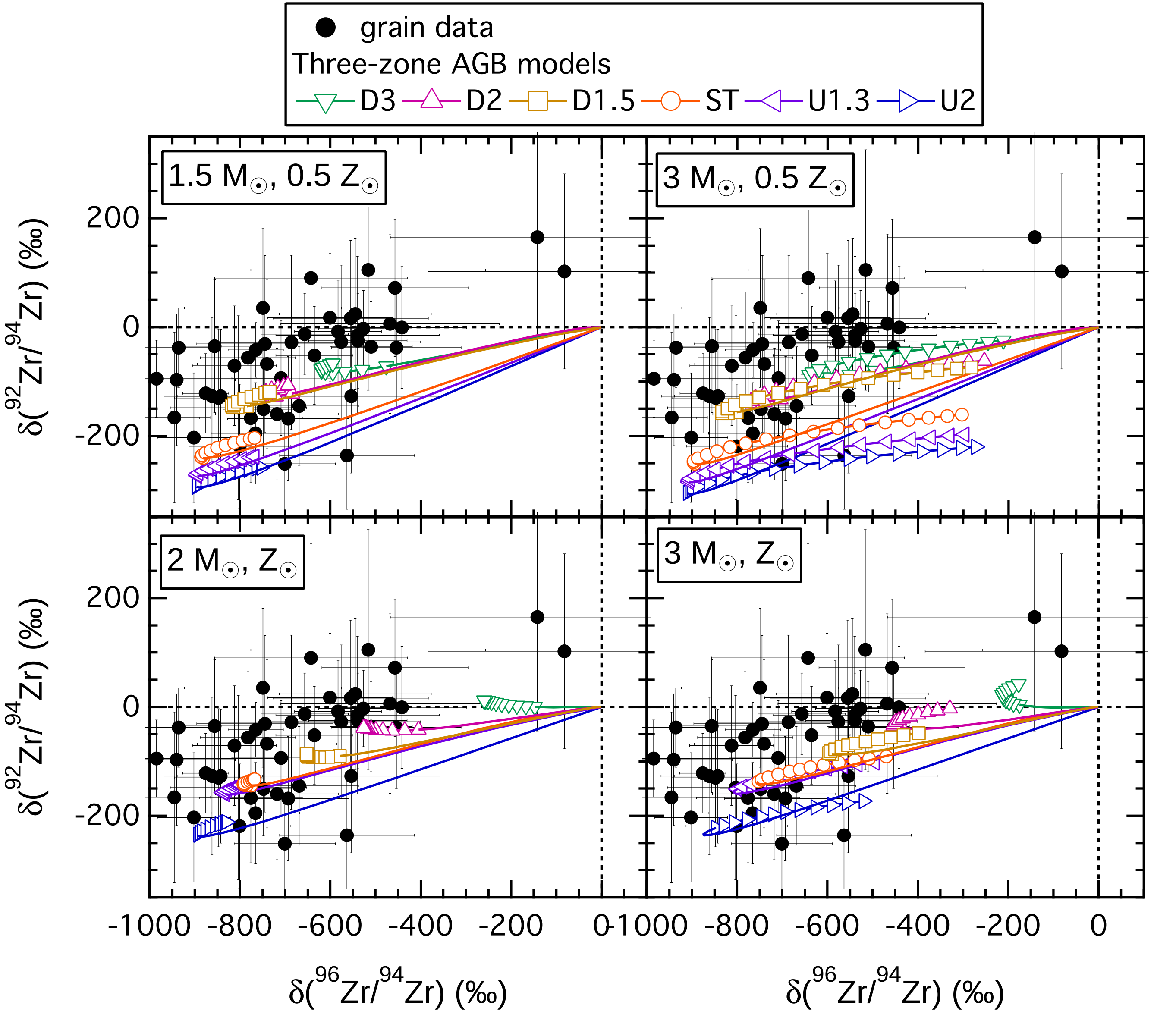

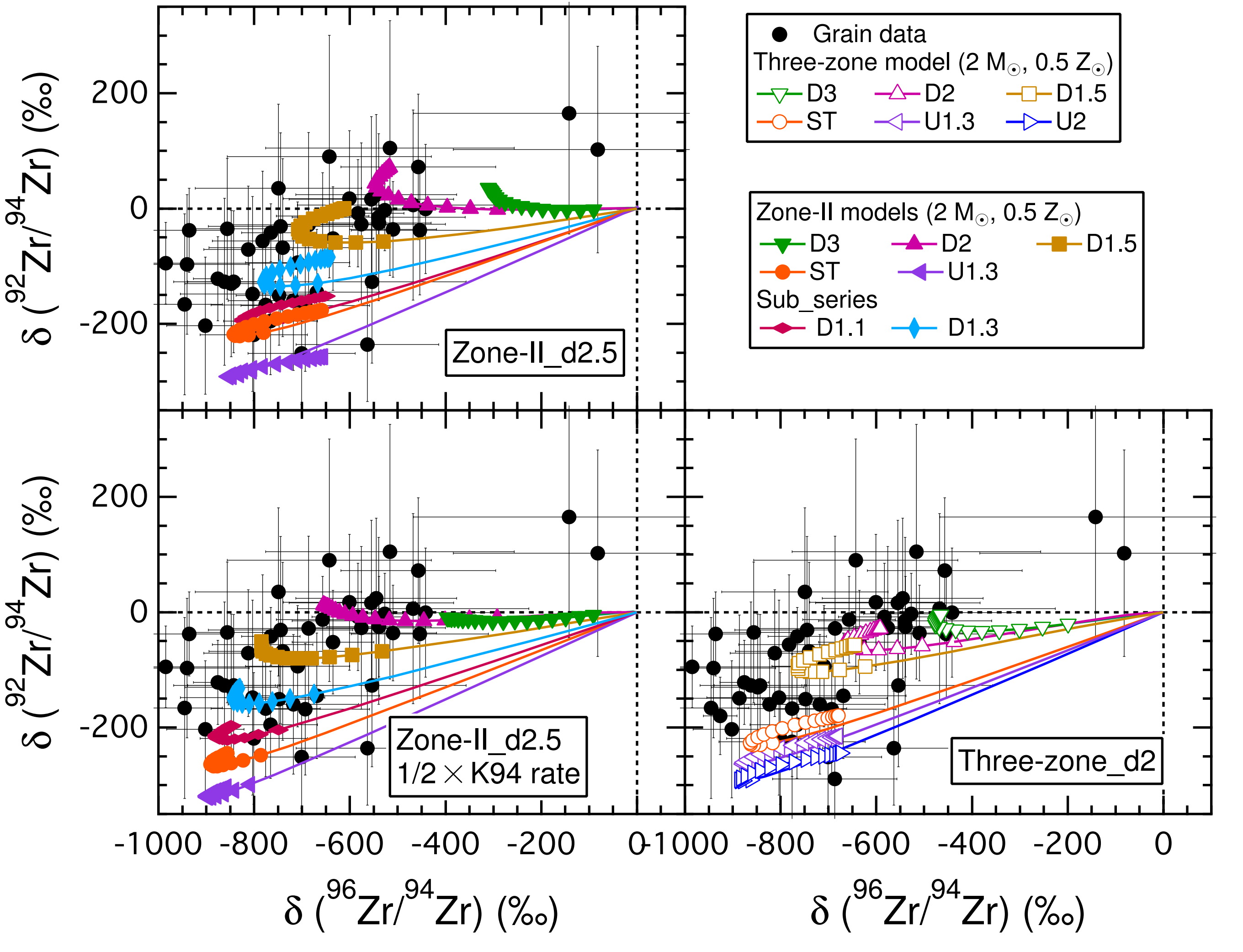

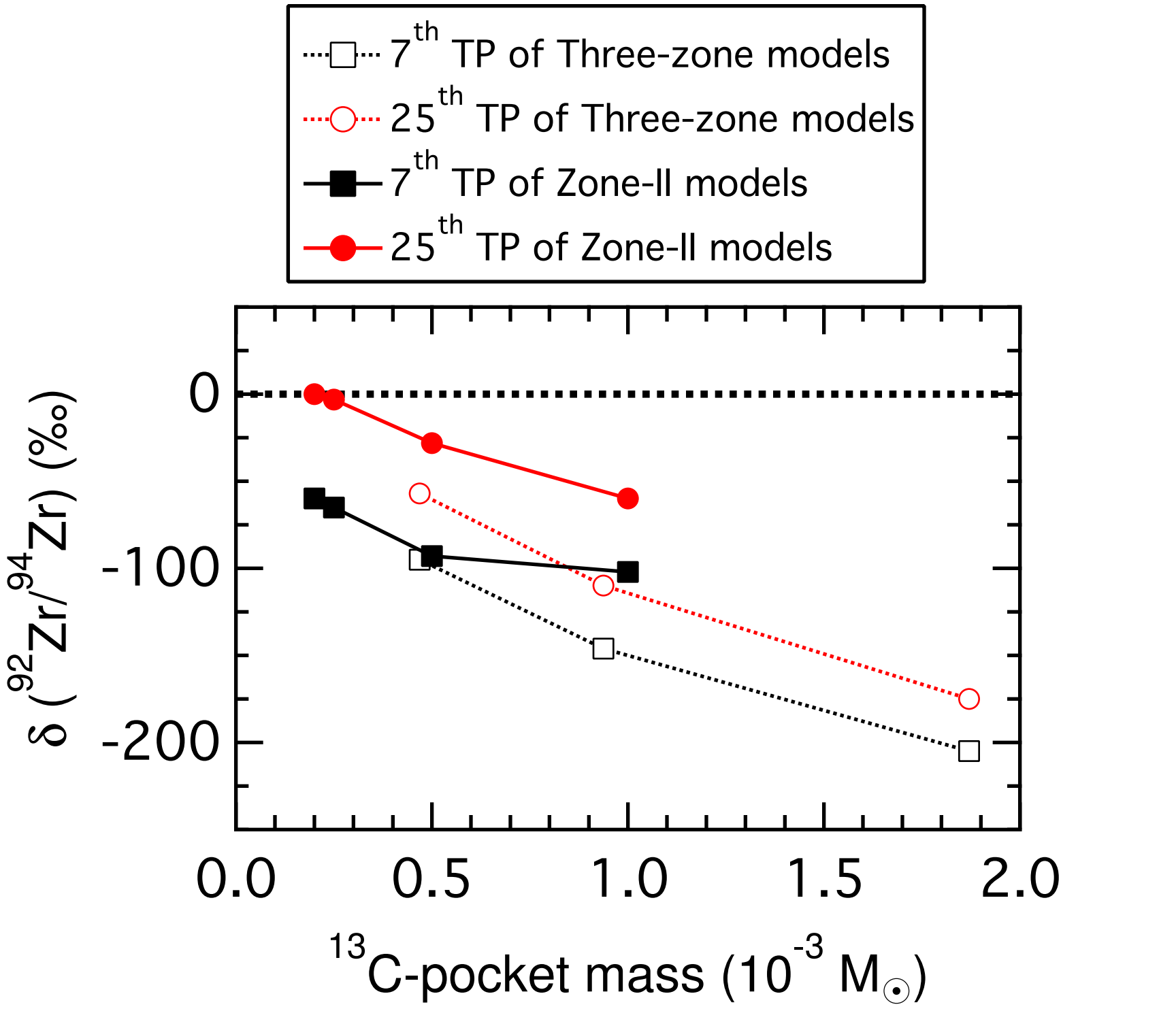

7th TP of Three-zone models
25th TP of Three-zone models
7th TP of Zone-II models
25th TP of Zone-II models
0
-100
-200
0.0
0.5
1.0
1.5
2.0
δ(92Zr/94Zr) (‰)
13C-pocket mass (10-3 M☉)

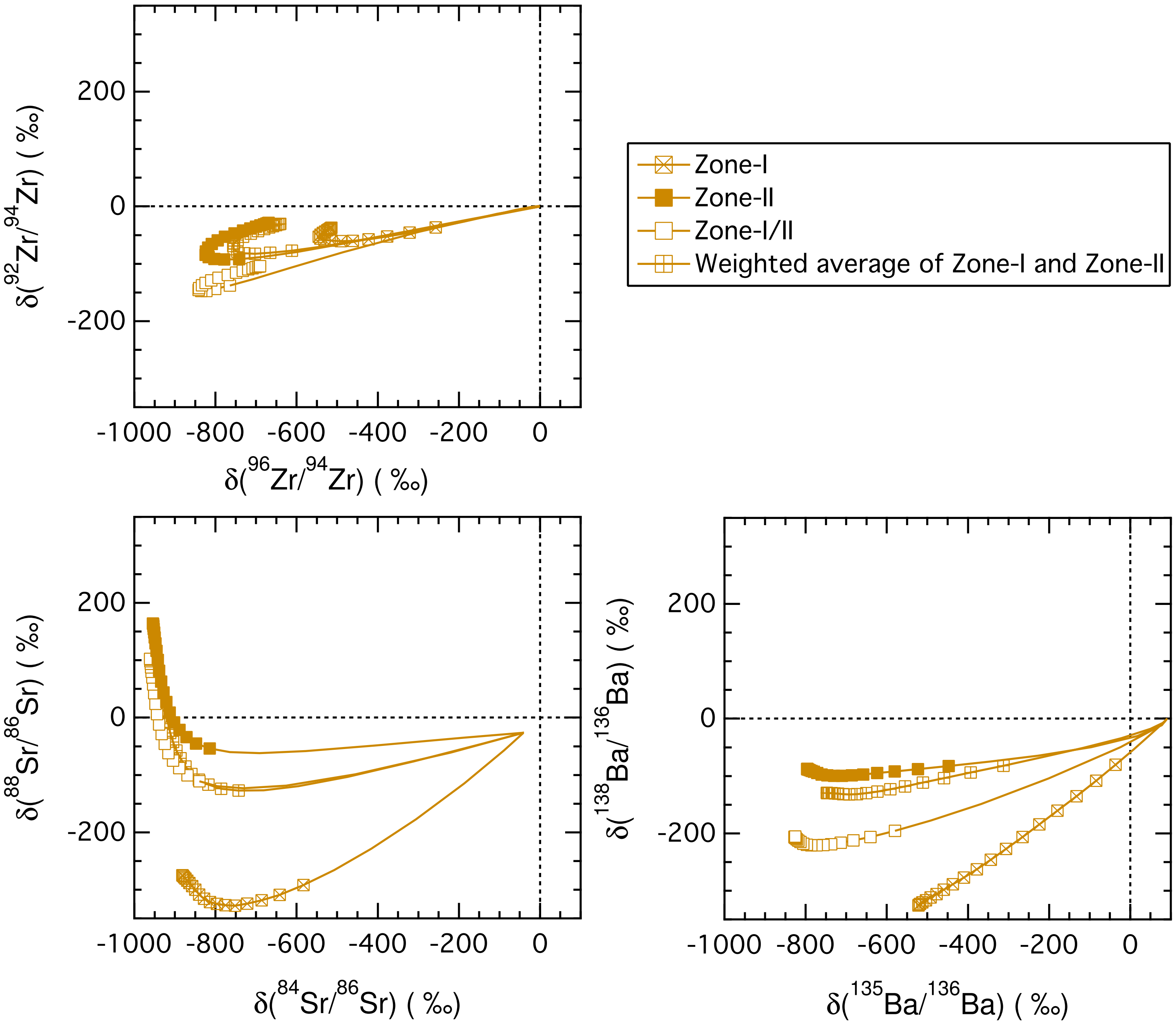